\documentclass[fleqn,twoside,twocolumn,nofootinbib]{revtex4} 
\usepackage{ujp} 
\begin{document}
\title[EFFECT OF UNIAXIAL STRESS]
{EFFECT OF UNIAXIAL STRESS ON LOW-FREQUENCY DISPERSION OF
DIELECTRIC CONSTANT\\ IN HIGH-RESISTIVITY GaSe CRYSTALS}%
\author{J.M. Stakhira}
\affiliation{Ivan Franko National University of L'viv}
\address{50, Dragomanov Str., L'viv 79005, Ukraine}
\email{flunt@electronics.wups.lviv.ua}
\author{O.Ye. Fl'unt}%
\affiliation{Ivan Franko National University of L'viv}%
\address{50, Dragomanov Str., L'viv 79005, Ukraine}%
\author{Ya.M. Fiyala}
\affiliation{Ivan Franko National University of L'viv}%
\address{50, Dragomanov Str., L'viv 79005, Ukraine}%
\email{flunt@electronics.wups.lviv.ua}%
\udk{}  \razd{}

\setcounter{page}{267}%
\maketitle

\begin{abstract}
Low-frequency dielectric spectra of high-resistivity GaSe layered
crystals have been studied on the samples clamped between two
insulating parallel plates at frequencies up to 100 kHz. The
measurements have been carried out at different uniaxial stresses
up to $2.4\times10^5$~Pa applied along the $c$-axis normal to crystal
layer's plane. It is revealed that the dielectric spectra of
high-resistivity GaSe layered crystals with insulating plates obey
a universal power law ${\sim}\omega^{n-1}$, where ${\omega}$ is the
angular frequency and $n\approx 0.8$, earlier observed on
high-resistivity GaSe crystals with indium-soldered contacts. The
same type of spectra on the crystals with different types of
contacts (insulating and ohmic) confirms the bulk character of
the observed polarization caused by hopping charge carriers. It is
shown that the frequency-dependent dielectric constant increases linearly
with the uniaxial stress characterized by
the coefficient
${\Delta}{\epsilon}/({\epsilon}{\Delta}{p})=8{\times}10^{-7}$~Pa$^{-1}$.
A slight increase of power $1-n$ with the stress is observed,
that leads to a stronger dielectric dispersion. The strong stress
dependence of the low-frequency dielectric constant in
high-resistivity GaSe crystals may be referred to the presence of
the formations of elementary dipoles, rotations of which
correspond to hops of localized charge carriers.
\end{abstract}

\section{Introduction}

Gallium selenide is a layered semiconductor with a band gap of 2~eV
at room temperature. Each layer of this crystal consists of four
sheets of like atoms in the sequence Se--Ga--Ga--Se, within which the
atoms are hold by strong covalent bonds \cite{struc72,struc73}. Weak
crystalline bonds between atoms, belonging to different neighbor
layers, and the possibility to stack the layers in different ways allow
the formation of a few polytypes. But gallium selenide crystals,
grown from a melt of the stoichiometric composition by the
Bridgman--Stockbarger method, usually belong to the $\epsilon$-polytype
\cite{struc82}, the hexagonal unit cell of which contains 8 atoms
and spans two layers in the direction normal to layer's plane.
The disposal of atoms inside a unit cell corresponds to the space
symmetry group~$\text{P-6m2}$~(187). The large crystallographic
anisotropy of GaSe crystals causes the singularities of mechanical,
electrical, and optical properties \cite{vlast1,vlast2,vlast3}. The high
electrical resistivity due to the wide band gap makes
gallium selenide attractive for studying the low-frequency
dielectric spectra and the effects appearing under condition of
the domination of displacement currents. From earlier studies
\cite{tezy}, it is known that the dielectric spectra of high-resistivity
GaSe crystals with resistivity over $10^5$~$\Omega$${\cdot}$cm are
characterized by the dispersion obeying the universal power law
\cite{Jon96}
\[ {\epsilon}_1(\omega)-j{\epsilon}_2(\omega)=\]
\begin{equation}
\label{univ}= B(j\omega)^{n-1}=B\left\{\sin{\frac{n\pi}2}-
j\cos{\frac{n\pi}2}\right\}\omega^{n-1},
\end{equation}

\noindent where ${\epsilon}_1(\omega)$ and ${\epsilon}_2(\omega)$
are real (in-phase) and imaginary (phase shifted by $\pi/2$)
frequency-dependent components of the complex relative dielectric
constant, respectively, $j$ is the imaginary unit, $\omega$ is the
angular frequency, and the exponent $1-n\approx0.2$. A dispersion of
this type, when the exponent $1-n<0.3$, but not too small, is
usually connected with the hopping transport of quasilocalized
electrical charge carriers \cite{Jon99,Jon96}. The localization of
charge carriers in GaSe crystals can be a result of the
non-controlled intercalation and a deviation of the chemical
composition from the stoichiometric one. In particular, according to
\cite{Maschke77,stacking2}, the local stacking faults of layers lead
to breaking down the translation symmetry of GaSe crystals along the
axis normal to layer's plane, that, in turn, leads to the
localization of charge carriers in the regions restricted by two
consequently displaced faults. Therefore, in most cases, layered
crystals and GaSe crystals, particularly, are characterized by a
high density of localized levels in the forbidden gap and in the
tails of bands of allowed energies. Recently, the low-frequency
dielectric spectra of GaSe crystals have been measured on samples
with nearly ohmic contacts fabricated by indium soldering on fresh
cleaved surfaces of the crystal \cite{tezy}. Studies of the
dielectric spectra of GaSe crystals with the use of blocking
contacts that do not allow the injection of excess charge carriers
into the crystal volume, as well as studying the influence of
external factors (particularly, a uniaxial pressure) on the
frequency-dependent dielectric permittivity, are of importance to
confirm the bulk character of observed polarization processes and to
obtain the information about singularities of the localization of
electrical charge carriers and the character of interaction between
them.

\begin{figure}
\includegraphics[width=\column]{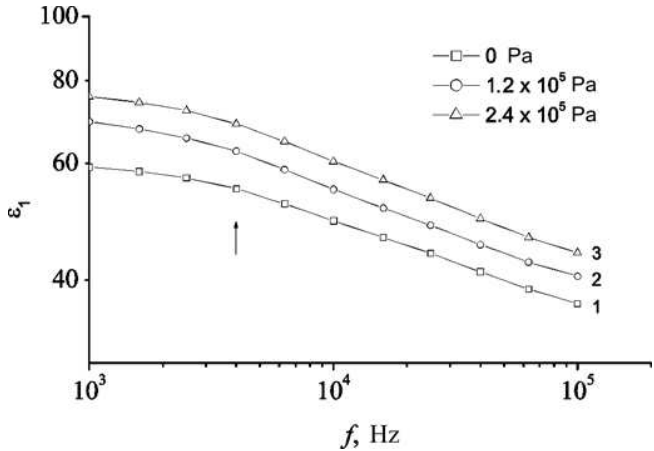}
\vskip-3mm\caption{Frequency dependencies of the relative dielectric
constant of a high-resistivity GaSe layered crystal at different
uniaxial stresses applied along the $c$-axis ({\it 1} -- 0~Pa, {\it
2} -- $1.2\times10^5$~Pa, {\it 3} -- $2.4\times10^5$~Pa)  }
\end{figure}

\section{Experimental}
This paper studies the low-frequency dielectric spectra of
high-resistivity GaSe crystals measured with the use of the plates
which are blocking (insulating) for the electrical charge carriers
under a uniaxial stress applied along the $c$-axis normal to
layer's plane. The disc-shaped ferroelectric capacitors with metal
plates ground from one side were used as blocking contacts
\cite{blocking1,blocking2}. The dielectric spectra of these plates
itself do not show a noticeable dispersion within the investigated
frequency range, and their capacity in the case of the squeezing
of the measuring cell without a sample is enough large (more than
50~pF) to create the conditions for prevailing the impedance of
the samples under study. Crystals for measurement were grown by
the Bridgman--Stockbarger method; samples were cleaved along
layers with a thickness of about 1~mm. The measurements of
dielectric spectra at frequencies up to 100~kHz have been carried
out, by transducing the complex capacity into a proportional ac
voltage. The transducer of the complex capacity fabricated on the
basis of an operational amplifier allows us to elucidate the
effect of the capacity of connecting cables on the measurement
results. In the process of measurement, the ac sinusoidal voltages
were consequently applied to the sample with different frequencies
with an approximately linear step on the log-scale in the set
frequency range with an rms voltage of 100~mV. The rms voltage on
the output of a transducer was measured with a digital voltmeter,
and the phase shift between input and output (proportional to the
current across the sample) signals with a digital phase-meter. The
preceding correction of the measuring equipment using the standard
elements allows us to avoid the constant bias. We note that the
impedance of dielectric plates decreases with increase in the
frequency according to the law ${\sim}1/\omega$ (in the case of
ideal capacity), and the impedance of high-resistivity GaSe
crystals in the region, where the displacement current prevails,
is proportional, as expected, to $\omega^{-s}$, where
$s\approx0.2.$ Therefore, at frequencies higher than some critical
value, the dielectric response of the sample under study will
dominate in the measuring spectra. We succeeded to observe this
domination of the dielectric spectrum of the sample under study
squeezed in a measuring cell with dielectric plates just on
high-resistivity GaSe crystals.

\section{Results and Discussion}
The frequency dependencies of the relative dielectric constant of a
high-resistivity GaSe layered crystal at various uniaxial stresses
are shown in Fig. 1. It corresponds to the real part of polarization
that changes in-phase with the applied sinusoidal electric field. It
can be observed on the frequencies higher than
${\sim}4{\times}10^3$~Hz, that is connected with the limiting effect
toward low frequencies of the capacities of dielectric plates of a
measuring cell which are connected schematically in-series. The
linear dependence of the dielectric constant on the frequency on the
log-log scale indicates that the spectra obey a universal power law
$\sim\omega^{-(1-n)}$, where $1-n\approx0.2$. The presence of the
dispersion in high-resistivity GaSe crystals, under condition of the
impossibility of the across current via a sample, is one more
confirmation of the bulk character of the polarization process under
study.

As can be seen from Fig. 2, the dispersive dielectric constant of
GaSe layered crystals linearly increases with the uniaxial
stress with a sufficiently high coefficient that equals, for example,
$3.5\times10^{-5}$~Pa at a frequency of 63~kHz. Since the spectra obey
the universal power law, it can be concluded, omitting a weak dependence
of the exponent $1-n$ on the pressure, that the ratio
of a relative change of the dielectric constant to a change of the stress
${\Delta}{\epsilon}/({\epsilon}{\Delta}{p})=
8{\times}10^{-7}$~Pa$^{-1}$ is independent of the frequency and can
characterize the sensitivity of the frequency-dependent dielectric
constant to a uniaxial stress in high-resistivity GaSe crystals.
The exponent $1-n$ can be obtained from the slope of
the frequency dependence ${\epsilon}_1(\omega)$ on the log-log scale or
from the ratio of the imaginary and real components of the dielectric
constant which is frequency-independent for a specific polarization process according to the universal power law \cite{Jon96}:
\begin{equation} \label{ratio}
\frac{{\epsilon}_2(\omega)}{{\epsilon}_1(\omega)}={\rm
ctg}\left(\frac{n\pi}2\right).
\end{equation}

In Table, we give the values of exponent $1-n$ which are
obtained from the ratios of the real and imaginary parts of
the dielectric constant (column 3) and from the slopes of the frequency
dependence of the dielectric constant on the frequency on the log-log scale
(column 4). Close values of $1-n$ obtained by two different
methods confirm that the response of a GaSe cystal under
study prevails in the high-frequency part
of the spectrum. The results shown in Table allow us to assert also that
the exponent $1-n$ weakly increases with the uniaxial stress.

\begin{table}[b]
\noindent\caption{Dependence of dielectric loss tangent and exponent
\boldmath$1-n$ of a GaSe crystal on the uniaxial stress at
a frequency of 40~kHz}\vskip3mm\tabcolsep16.0pt

\noindent{\footnotesize\begin{tabular}{c c c c }
 \hline \multicolumn{1}{c}
{\rule{0pt}{9pt}$p$, Pa} & \multicolumn{1}{|c}{${\rm tg}\delta$,
40~kHz}& \multicolumn{1}{|c}{$1-n$}&
\multicolumn{1}{|c}{$1-n$}\\%
\hline%
0 & 0.21 & 0.132 & 0.129\\%
$1.2\times10^5$ & 0.223 & 0.139 & 0.139\\%
$2.4\times10^5$ & 0.237 & 0.148 & 0.144\\%
\hline
\end{tabular}}
\end{table}

The measured ratios of a relative change of the
frequency-dependent dielectric constant to a uniaxial stress in
GaSe layered crystals significantly exceed the piezoconductivity
coefficients for most known semiconducting materials, which is
explained by a transformation of the energy band spectra of
crystals and a change of the effective mass of charge carriers
\cite{piezo}. Further, let us estimate a possible change of the
dielectric constant within the model of hops of localized charge
carriers in rigid double potential wells which corresponds to the
absence of interaction between charges. For example, the hopping
frequency depending on the tunneling factor can be written as
$f=f_0\exp(-2{\alpha}R)$, where $f_0$ is the attempt frequency,
$\alpha$ is the wave function decay constant, and $R$ is the
distance between localization centers. A change of the distance
between two centers by ${\Delta}R$ leads to a relative change of
the hopping frequency
${\Delta}f/f=-2{\alpha}{\Delta}R=ln(f/f_0)({\Delta}p/B_{3})$,
where ${B}_{3}$ is the elasticity modulus of a GaSe crystal along
the $c$-axis. The component of the elasticity modulus
characterizing a change of interlayer distances in GaSe layered
crystals \cite{interlayer}, which is significantly less than that
connected with a change of the distances between atoms inside
layers, can be considered as ${B}_{3}$. But even in this case, the
relative change of the hopping frequency at the applied stress
$2.4{\times}10^5$~Pa is about $10^{-4}$, which leads to an even
less relative change of the dielectric constant. Therefore, the
model of independent hops of charge carriers in rigid double
potential wells, for which the distance between energy minima
changes proportionally to a relative change of interlayer
distances, is also incapable to explain the observed change of the
frequency-dependent dielectric constant of crystals under the
action of a uniaxial stress.

\begin{figure}
\includegraphics[width=7.5cm]{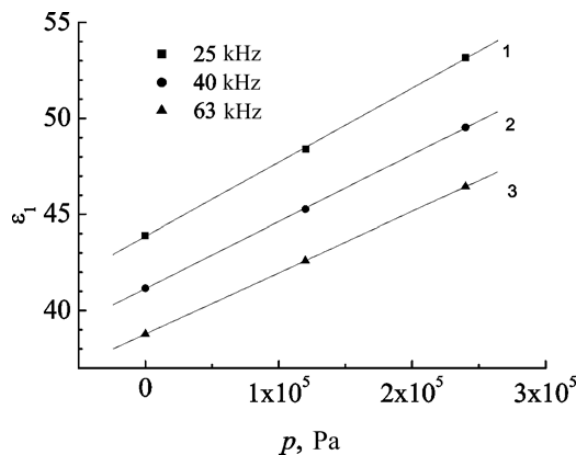}
\vskip-3mm\caption{Dependencies of the relative dielectric constant
of a high-resistivity GaSe layered crystal on the uniaxial stress
applied along the $c$-axis measured at various frequencies ({\it 1}
-- 25~kHz, {\it 2} -- 40~kHz, {\it 3} -- 63~kHz)  }
\end{figure}

But, according to the modern conceptions, the polarization in
condensed matter cannot be considered as a rotation of independent
dipoles, each of which changes the orientation in rigid double
potential wells or wells with more minima \cite{Jon96,Dissado}. A
change of the position of each localized charge carrier equivalent
to the rotation of an elementary electric dipole leads to a change
of the dispositions of many other neighbor localized charges within
the region of a certain radius. So, the experimentally measured
macroscopic parameters of polarized condensed media characterize not
the individual dipoles, but some formations on the basis of
elementary dipoles. At~present, the behavior and dimensions of these
formations are known very little. They are characterized by their
own effective dipolar moments and relaxation times which differ from
those of elementary dipoles and cause, in most cases,  a
distribution of relaxation times in a wide interval. A significant
sensitivity of the dielectric constant of GaSe crystals to a
mechanical stress can be explained by the presence of the formations
of elementary dipoles, whose rotations are equivalent to localized
charge carrier hops. As a result, each elementary localized charge
does not stay not in some mean electric field, but undergoes the
action of a local electric field formed by a certain arrangement of
surrounding electric charges, which is established in the process of
creation of dipole's formations at a certain time moment. The
establishment of a distribution of local electric fields and the
creation of formations of elementary dipoles are random processes
strongly dependent on previous step-by-step changes in the dipolar
system and thereby very sensitive to defects of the crystal
structure and different external factors, particularly a uniaxial
stress. An insignificant influence of some external factors can
decline the preferred positions from one side to another one for
some individual dipoles with close values of probabilities to be in
two different directions. As a result of the subsequent processes of
reorientation of the neighboring dipoles, this can cause a
significant change of parameters of the formations of individual
dipoles, which will reveal, in our case, in changes of the
low-frequency dielectric constant under a uniaxial stress.

\section{Conclusions}

Our investigations of the low-frequency dielectric spectra of
high-resistivity GaSe crystals clamped between dielectric plates of
a flat capacitor demonstrated the dominance of the dielectric
spectrum of the $\omega^{n-1}$ type, where $\omega$ is the angular
frequency and the exponent $n~{\approx}~0.8$, which is
characteristic for the systems with hopping motion of charge
carriers. It is established that the disperse permittivity of
high-resistivity GaSe crystals linearly depends on the uniaxial
stress and can be characterized by the frequency-independent
coefficient ${\Delta}{\epsilon}/({\epsilon}{\Delta}{p})=
8{\times}10^{-7}$~Pa$^{-1}.$ The exponent $1-n$ insignificantly
grows with increase in the uniaxial stress. The revealed
considerable dependence of the low-frequency permittivity of
high-resistivity GaSe crystals on the uniaxial stress is related to
the formation of ensembles of elementary dipoles in the crystal,
whose rotations are equivalent to hops of localized charge carriers.
The established regularities of the dependence of the dispersive
permittivity on the uniaxial stress can be used for studying the
peculiarities of the formation of dipole ensembles in solids, as
well as for creating the uniaxial stress sensors based on new
physical phenomena.

\rezume{ВПЛИВ ОДНОВІСНОГО ТИСКУ НА НИЗЬКОЧАСТОТНУ ДИСПЕРСІЮ
ДІЕЛЕКТРИЧНОЇ ПРОНИКНОСТІ \\У ВИСОКООМНИХ КРИСТАЛАХ GaSe} {Й.М.
Стахіра, О.Є. Флюнт, Я.М. Фіяла} {Проведено дослідження
низькочастотної діелектричної проникності високоомних кристалів GaSe
на частотах до 100~кГц з використанням блокуючих для носіїв
електричного заряду (ізолюючих) контактів. Вимірювання проводили при
прикладанні до зразка невеликого одновісного тиску в межах до
$2,4\cdot 10^5$~Па вздовж осі $c$, нормальної до площини шарів
кристала. Встановлено, що діелектричний спектр високоомних кристалів
GaSe з блокуючими електродами підлягає універсальному степеневому
закону ${\sim}\omega^{n-1}$, де ${\omega}$ -- кутова частота,
$n\approx 0,8$, який раніше спостерігали на високоомних зразках з
контактами з наплавленого індію. Однакова форма діелектричного
спектра на кристалах з різними типами контактів (омічними та
блокуючими) підтверджує об'ємний характер спостережуваного явища
поляризації, яке пов'язується зі стрибкоподібним переміщенням
квазілокалізованих носіїв електричного заряду. Встановлено, що
діелектрична проникність лінійно зростає з величиною прикладеного
одновісного тиску з коефіцієнтом
${\Delta}{\epsilon}/({\epsilon}{\Delta}{p})= 8{\cdot
}10^{-7}$~Па$^{-1}$. Спостерігається незначне збільшення показника
степеня $1-n$ при збільшенні тиску, що приводить до посилення
дисперсії діелектричної проникності. Значна залежність
низькочастотної діелектричної проникності від одновісного тиску в
високоомних кристалах GaSe пов'язується з формуванням утворень
диполів, обертання яких\linebreak еквівалентні стрибкам
локалізованих носіїв електричного заряду.}

\end{document}